\documentclass{iaus}
\usepackage{graphicx}

\newcommand{\ltapprox}{\raisebox{-0.5ex}{$\,\stackrel{<}{\scriptstyle\sim}\,$}}
\newcommand{\gtapprox}{\raisebox{-0.5ex}{$\,\stackrel{>}{\scriptstyle\sim}\,$}}
\newcommand{\lya}{\ifmmode {\rm Ly}\alpha \else Ly$\alpha$\fi}
\def\msun{$M_\odot$}
\def\lya{Lyman-$\alpha$}

\def\msun{\ifmmode M_{\odot} \else M$_{\odot}$\fi}
\def\zsun{\ifmmode Z_{\odot} \else Z$_{\odot}$\fi}
\def\lsun{\ifmmode L_{\odot} \else L$_{\odot}$\fi}

\title[Very High-Redshift Lensed Galaxies]{Very High-Redshift Lensed Galaxies}

\author[Pell\'o et al.]%
{R. Pell\'o$^1$,
D. Schaerer$^{2,1}$, J. Richard$^1$, \break
J.- F. Le Borgne$^1$, 
\and J.- P. Kneib$^{3,4}$}

\affiliation{$^1$Laboratoire d'Astrophysique de l'Observatoire
Midi-Pyr\'en\'ees, UMR5572, 14 Av. Edouard Belin, F-31400 Toulouse,
France, email:  roser, leborgne, jrichard@ast.obs-mip.fr \\[\affilskip]
$^2$Observatoire de Gen\`eve,
51, Ch. des Maillettes, CH-1290 Sauverny, Switzerland,
email:daniel.schaerer@obs.unige.ch \\[\affilskip]
$^3$Caltech Astronomy, MC105-24, Pasadena, CA 91125,
USA, email: kneib@caltech.edu \\[\affilskip]
$^4$Laboratoire d'Astrophysique de Marseille, Traverse du Siphon,
B.P.8, 13376 Marseille Cedex 12, France}

\pubyear{2004}
\volume{225}
\pagerange{1--8}
\date{?? and in revised form ??}
\jname{Impact of Gravitational Lensing on Cosmology}
\editors{Mellier, Y. \& Meylan,G. eds.}
\begin{document}

\maketitle

\begin{abstract}
We review in this paper the main results recently obtained on the
identification and study of very high-z galaxies using lensing
clusters as natural gravitational telescopes. We present in detail our pilot
survey with ISAAC/VLT, aimed at the detection of z$>7$ sources.  
Evolutionary synthesis models for extremely metal-poor and PopIII 
starbursts have been used to derive the observational properties
expected for these high-$z$ galaxies, such as expected magnitudes and
colors, line fluxes for the main emission lines, etc. These models have
allowed to define fairly robust selection criteria to
find z$\sim 7-10$ galaxies based on broad-band near-IR
photometry in combination with the traditional Lyman
drop-out technique. The first results issued from our photometric and
spectroscopic survey are discussed, in particular the preliminary
confirmation rate, and the global properties of our high-z candidates,
including the latest results on the possible z=10.0 candidate
A1835-1916. The search efficiency should be significantly improved
by the future near-IR multi-object ground-based and space
facilities. However, strong lensing clusters remain a factor of $\sim
5-10$ more efficient than blank fields in the z $\sim 7-11 $ domain,
within the FOV of a few arcminutes around the cluster core,
for the typical depth required for this survey project.

\end{abstract}

\firstsection 
\section{Introduction}\label{intro}

Considerable advances have been made in the exploration of the high-z
Universe during the last decade, starting with the discovery and
detailed studies of redshift $z\sim 3$ galaxies (Lyman break
galaxies, LBGs), mostly from the pioneering work of Steidel
and collaborators (cf.\ Steidel et al.\ 2003, Shapley et al.\ 2003),
the $z \sim$ 4--5 galaxies found from different deep
multi-wavelength surveys, to galaxies at $z \sim$ 6--7, close to the
end of the reionisation epoch of the Universe (e.g. Kodaira et al.\ 2002, 
Hu et al.\ 2002, Cuby et al.\ 2003, Kneib et al. 2004, Stanway et 
al.\ 2004, Bouwens et al.\ 2004). 
To extend the present
searches beyond $z\ge$ 6.5 and back to ages where the Universe was
being re-ionized (cf.\ Fan et al.\ 2002), it is mandatory to move into
the near-IR bands.
According to WMAP results, the first
generation of stars may exist at $z \sim$ 14--20
(Bennet et al. 2003; Spergel et al. 2003). 

    At z$>7$, the most relevant signatures are expected in the near-IR
$\lambda \gtapprox 1 \mu$m window. Most feasibility studies during the last
years have been motivated by JWST, which should be able to observe
galaxies up to redshifts of the order of $z \sim$ 20-30. 
However, the delay of JWST and the availability of well suited 
near-IR facilities in ground-based 8-10m telescopes hasten the
development of observational projects targeting $z \gtapprox 7$ sources.
The first studies on the physical properties of these sources
can be started nowadays with instruments such as ISAAC/VLT, and pursued
with improved efficiency using the future multi-object spectrographs
to come, such as EMIR at GTC ($\sim$ 2006), or
KMOS for the second generation of VLT instruments ($\sim$
2009). In other words, the efficiency of this research is closely
related to the instrumental developments in the near-IR domain. 

We present here the first  
results from our ongoing pilot project with ISAAC/VLT, aimed at the
detection of $z>7$ sources through lensing clusters used 
as Gravitational Telescopes (GTs). Our goal is to build up the first
spectroscopic sample of galaxies at $z>7$ starting with GTs, in
preparation of future massive surveys in the field. Such sample should
allow to start a detailed study on the physical properties of these
objects (SFR, extinction, metallicity, IMF, ...), and also start
constraining their global properties (number counts, luminosity
function, clustering properties, ...). 

 The plan of the paper is as follows. We review in
section~\ref{identification} the techniques and main results recently
obtained on the identification of high-z galaxies in the optical bands,
both in the field and using lensing clusters. In Sect.~\ref{z7-10} we summarize 
the observational properties expected for galaxies at 7 $\ltapprox$ z
 $\ltapprox$ 11. Our pilot survey with ISAAC/VLT, aimed at the
 detection of z$>7$ sources, is presented in details in
 Sect.~\ref{VLTsurvey}. The first results issued from our photometric and
spectroscopic survey are discussed in this section, in particular the preliminary
confirmation rate, and the global properties of our high-z candidates,
including the latest results on the possible $z=10.0$ candidate
A1835-1916. A discussion on the preliminary results and constraints
derived from the 7$\le$ z $\le$ 10 photometric candidates is given in
Sect.~\ref{discussion}, together with future prospects. Throughout
this paper, we adopt the cosmological parameters:
 $\Omega_{\Lambda}=0.7$, $\Omega_{m}=0.3$ 
and $H_{0}=70$ km s$^{-1}$ Mpc$^{-1}$.

\section{Looking for $z \sim 5-7$ galaxies in the optical bands}\label{identification}


Different field surveys in the optical bands have been successful in
the systematic search for galaxies close to the reionisation
epoch. These surveys are mainly based on 2 different techniques, which
can easily  be exploited in the near-IR bands: narrowband (NB) imaging
surveys aimed at detecting \lya \ emission, and broad-band multi-color
surveys using photometric redshifts at some level (see a review by
Spinrad, 2003). 

The pioneering Large Area Lyman Alpha Survey (LALA, Rhoads \& Malhotra 2001,
Rhoads et al. 2003) used NB imaging together with $BVR$
broad-band photometry to target strong \lya \ emission
at $z\sim5.7$ in a KPNO + CTIO  $36' \times 36'$ field of
view. The spectroscopic follow up at LRIS/Keck confirmed 3/4
candidates, with important line fluxes (a few $\times 10^{-17}$ erg
cm$^{-2}$ s$^{-1}$) and $\sigma_v\ge 200$ km/s.
Rhoads et al. (2004) have recently published new results at higher
redshifts: a z=6.535 galaxy, representing 
1/3 of candidates confirmed, with restframe $\sigma_v\ge 40$ km/s
and \lya \ flux $2 \times 10^{-17}$ erg cm$^{-2}$ s$^{-1}$.
Other NB surveys in the field have also provided
interesting galaxy samples in the z$\sim 5-7$ interval: the z=6.17
galaxy found by Cuby et al. (2003) in the CFHT-VIRMOS Deep Survey; the 
two $z\sim 6.6$ galaxies detected by Kodaira et al. (2003), and the
large sample of \lya \ emitters at $z\sim5.7$ by Hu et
al.\ (2004). In the later case, the authors used a combination of
NB imaging at 8150\AA \ (SuprimeCam) and broad-band photometry
in the optical bands to select candidates for a subsequent
spectroscopic follow up with DEIMOS/Keck. Their confirmation rate is
very high: 18 sources out of 26 candidates are confirmed at high-z in
a $34' \times 27'$ field of view, thus leading to 0.03
sources/arcmin$^2$ and redshift bin $\delta z=0.1$. All these sources
have important \lya \ fluxes (a few $\times 10^{-17}$ erg
cm$^{-2}$ s$^{-1}$), and display broad \lya \ lines ($\sigma_v \sim
200$ km/s).


The broad-band photometric selection of high-z sources is usually an
extrapolation of the drop-out LBGs technique to higher redshifts, with
subsequent spectroscopic identification of the selected candidates. 
When the photometric spectral coverage is wide enough, standard
photometric redshifts can be derived through
SED fitting procedures. An excellent example of this technique is the
sample at $z \sim 4.5-6.6$ selected by Lehnert \& Bremer
(2003) in a 44 arcmin$^2$ field, using an $Riz$ color criterium. 6
out of 12 "good" candidates were spectroscopically confirmed with
FORS/VLT, thus leading to a number density for these sources which is
close to LBGs at lower redshifts.
Also Bouwens et al.\ (2003) and Stanway et al.\ (2004) 
have recently derived the $z\sim 6$ star
formation density from $I$-dropouts candidates at this redshift
selected on ACS/GTO HST fields. 

Gravitational magnification by lensing clusters has proven highly
successful in identifying a large fraction of the most distant
galaxies known today. 
The systematic search for high-z targets around the critical lines
allowed Ellis et al. (2001) to find a lensed galaxy at z=5.58 behind
A2218, spectroscopically confirmed at Keck. 
This source is a faint
compact galaxy, dominated by its \lya \ emission, the stellar continuum
beeing negligible in this case. 
Another interesting object recently discovered by Hu et al. (2002) is
a lensed galaxy at z=6.56 behind the cluster A370. The
authors used NB imaging at 9152 \AA \, obtained with LRIS on
the Keck telescope, to identify the image by the equivalent width 
of the line emission and the absence of flux at lower wavelengths. As
it happens with sources identified from NB imaging, the flux
measured in the line is relatively high:$ 2.7\times 
10^{-17}$ erg cm$^{-2}$ s$^{-1}$. The magnification factor is
moderate, only 4.5, the source beeing located more than 1'
from the cluster center. 

A deep and ``blind'' spectroscopic survey for high-$z$ \lya \ emitting
sources was performed by Santos et al.\ (2004), around the
critical lines in 9 intermediate-$z$ lensing clusters. They
targeted candidates at $4.5 \le z \le 6.7$, magnified by a factor of
at least 10. Eleven of such candidates were later confirmed at $2.2 \le z \le
5.6$ with LRIS long-slit and ESI spectra at Keck, taking advantage
from the higher resolution to resolve the [OII]$\lambda$3727 doublet
or \lya \ asymmetric profiles to discriminate between low and high-$z$
sources. Thanks to the lensing magnification, these authors derived
the \lya \ Luminosity Function to unprecedented depths (see also
J. Richard's poster, this conference).

In a recent paper, Kneib et al.\ (2004) discuss the 
identification of a possible $z\sim7$ compact galaxy behind A2218. 
This is a faint triply imaged galaxy identified from deep
$z$-band imaging with ACS/HST. Although there is no emission line
detected in this object, both photometric and lensing
modelling considerations provide a fair determination of its
redshift. Regardless of the precise redshift, the
multi-wavelength data available for this object already provide an
interesting information on its nature (see also D. Schaerer's talk and
J. Richard's poster, this conference). The important point here is
that GTs allow to define different samples of high-$z$ sources, using
the same criteria as for field surveys (NB or broad-band
imaging), but also some specific criteria based on lens-modelling
considerations (systematic searches around the critical lines). All of
them have a common point: the final samples will consist of sources
intrinsically fainter than the field ones. The search efficiency as
compared to blank fiends depends on two opposite trends:
gravitational magnification (allowing to increase the number of faint
sources, and thus the total number of sources at the end), and
dilution due to the reduction of the effective area surveyed. As shown
below, strong lensing clusters remain a useful tool for $z \ge 7$
studies. 

\section{Expected properties of galaxies at z$\ge 7$}\label{z7-10}

\begin{figure}
\includegraphics[height=3.5in,width=5in,angle=0]{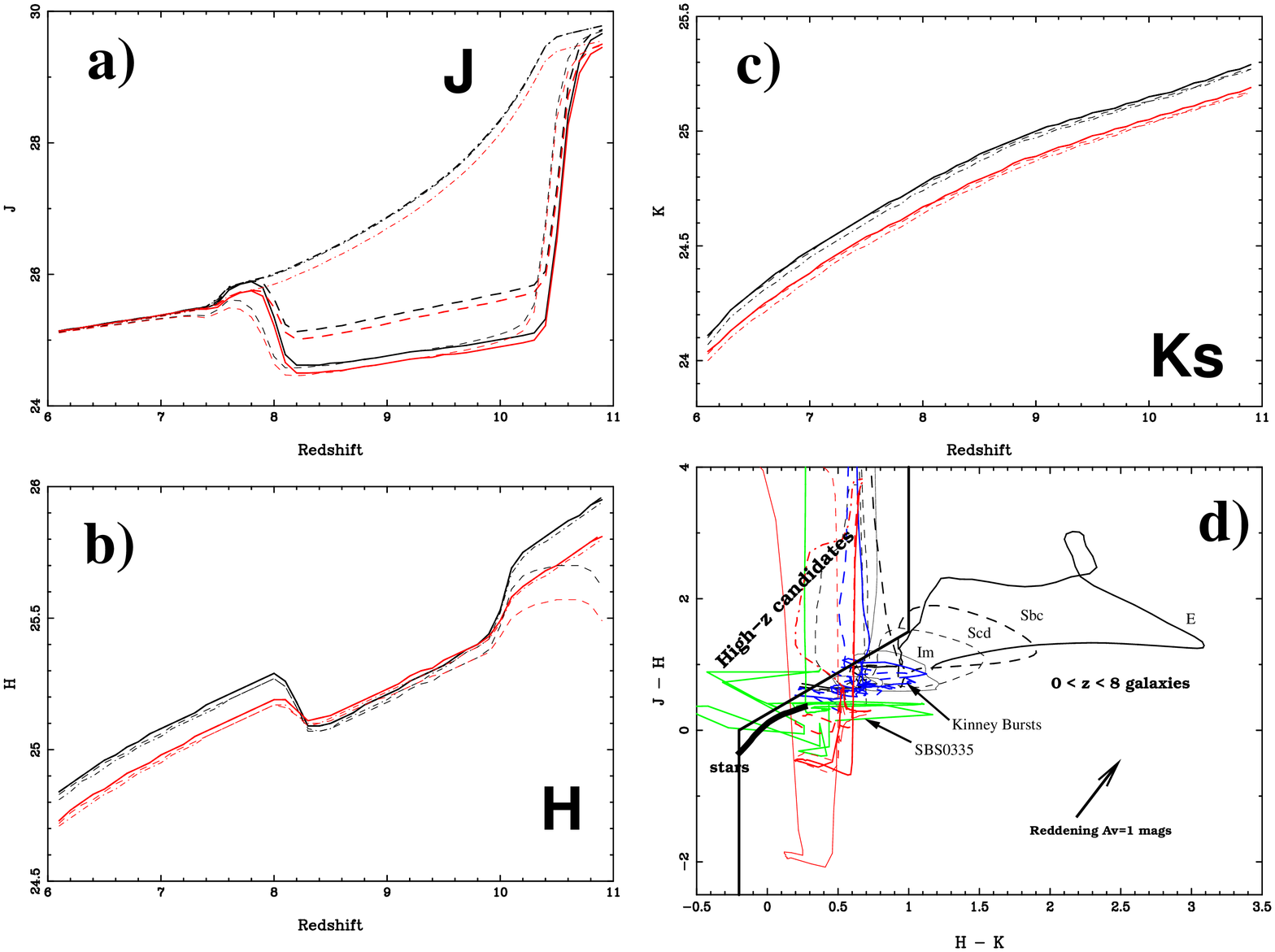}
  \caption{(\textit{a}) $J$(Vega) magnitude as a function of redshift
for a top-heavy IMF, for a fiducial stellar halo of $10^7$
M$_{\odot}$. The values corresponding to a normal Salpeter IMF are
about 2 magnitudes fainter over all the redshift interval.
Black and red (grey) lines correspond respectively to burst ages
$10^4$ and $10^6$ yrs. Various models for Pop III objects are
presented, for different fractions of the Lyman $\alpha$ emission
entering the integration aperture: 0\% (thick dot-dashed line),
50\% (thick dashed line), and 100\% (thick solid line). Thin
dot-dashed lines correspond to a self-consistent
extended Lyman$\alpha$ halo emission (Loeb \& Rybicki 1999), whereas
thin dashed lines display the same model with 10\% of Lyman$\alpha$
emission entering the integration aperture. (\textit{b}) Same as a) for $H$(Vega)
magnitude versus redshift. (\textit{c} Same as a) for $Ks$(Vega)
magnitude versus redshift.
(\textit{d}) $J-H$ versus $H-Ks$ color-color diagram (Vega system) showing the position
expected for different objects over the interval of $z \sim$ 0 to 11.
The position of stars and normal galaxies up to z $\le$ 8
are shown, as well as the shift direction induced by reddening A$_V=1$
magnitude. Various models for Pop III objects are presented, with
different fractions of Lyman~$\alpha$ emission entering the
integration aperture: 100\% (thick dot-dashed line), 50\% (thick dashed
line), and 0\% (thick solid line). Thin solid and dashed lines
correspond to a core (non-resolved) source, whith extended
Lyman~$\alpha$ halo, for 50\% and 10\% of Lyman~$\alpha$ emission
entering the integration aperture. 
The location of starbursts templates (SB1 and SB2, from Kinney et al. 1996),
and the low metallicity galaxy SBS0335-052 are also given for
comparison. All star-forming models enter the high-$z$
candidate region at z$ge$ 8.}\label{fig:cc_theoriques}
\end{figure}

   Our project is to start looking for $z \sim$ 7--11 galaxies. At such
redshifts, the most relevant signatures are expected in the 
$\lambda \gtapprox 1 \mu$m domain. We have used the evolutionary
synthesis models by Schaerer (2002, 2003) for Population III and
extremely metal deficient starbursts to derive the expected magnitudes and
colors of galaxies in this redshift domain. For genuine
PopIII sources, nebular continuous emission dominates the ZAMS 
spectra at $\lambda\ge 1400$ \AA \ and strong emission lines are
present (Lyman~$\alpha$, HeII$\lambda$1640, HeII$\lambda$3203,
...). The IMF could be dominated by very massive stars, up to $\sim
1000 M_{\odot}$, with a possible lack of low-mass stars 
(top-heavy IMF, Abel et al. 1998, Bromm et al. 1999). 
An important issue for our project is the photometric selection of
candidates allowing a subsequent successful spectroscopic follow
up. 

Simulations have
been done to define the observing strategy to target $z>7$ sources.
We consider a fiducial stellar mass halo of $10^7$ M$_{\odot}$,
corresponding to a collapsing DM halo of $2 \times 10^8 M_{\odot}$,
thus typically to $\sim$ 1.5 to 2 $\sigma$ fluctuations between z=5 and 10
(e.g. Loeb \& Barkana 2001). Magnitudes are to be
rescaled according to this value for other mass halos. These sources
are espected to be unresolved on 0.3'' scales. The reionization
redshift is assumed to be $z \sim 6$.  Lyman series
troughs (Haiman \& Loeb 1999), and Lyman forest following the
prescription of Madau (1995) are included. Simulations accounting
for an extended \lya \  halo (cf. Loeb \& Rybicki 1999) have also been
computed, together with simple assumptions for the Lyman $\alpha$
emission line with arbitrary different fractions of the emission flux
entering the integration aperture. Two extreme assumptions are
considered here for the IMF, either a standard Salpeter IMF, with
stars forming between 1 and 100 $M_{\odot}$, or a top-heavy Salpeter IMF, with
stellar masses ranging between 50 and 500 $M_{\odot}$.
Early versions of these simulations were presented in
Schaerer \& Pell\'o (2001) and Pell\'o \& Schaerer (2002), as
feasibility studies for EMIR/GTC and ISAAC-KMOS/VLT.

Genuine $z>7$ sources are optical dropouts.
Fig.~\ref{fig:cc_theoriques} displays the $J-H$ versus $H-Ks$
diagram for different models compared to the 
location of stars and normal galaxies at $0 \le z \le
8$. We used the set of empirical SEDs compiled by Coleman, Wu
and Weedman (1980) to represent the local population
of galaxies, with spectra extended to wavelengths $\lambda \le
1400$\,\AA\ and $\lambda\ge 10000$\,\AA\ using the equivalent
spectra from the GISSEL library for solar
metallicity (the Bruzual \& Charlot 1993). We also
included the starbursts templates SB1 and SB2, from Kinney et
al.\ (1996), and the low metallicity galaxy SBS0335-052.
The shift in this diagram corresponding to an intrinsic reddening of
A$_V$=1, with the reddening law by Calzetti et al. (2000), is shown
by an arrow. 
As shown in
Fig.~\ref{fig:cc_theoriques}, the $J-H$ versus $H-Ks$ color-color diagram of
optical dropouts is particularly well suited to identify galaxy
candidates at $8 \le z \le 11$. At redshifts
above $z\sim10$, galaxies are expected to be non-detected in the $J$-band.  
At $6 \le z \le 9$, the same photometric selection can be
performed including the $Z$ (0.9 $\mu$m) and $SZ$ (1.1 $\mu$m) filters.

   An important issue is the photometric depth needed to detect
typical stellar haloes up to a given mass.
According to our simulations, the predicted (Vega) magnitudes
for the reference stellar halo,
assuming a top-heavy Salpeter IMF, range between
$\sim$ 24.5 and 26.5 in $J$, $\sim$ 24.5 and 25.5 in $H$, and
$\sim$ 24 to 25 in $Ks$, depending on models,
within the $z \sim$ 7--10 interval
(Fig.~\ref{fig:cc_theoriques}). These values are about 2 magnitudes
fainter for a standard Salpeter IMF. If we intend to detect typical stellar
haloes up to a few $10^7$ M$_{\odot}$, and a significant fraction of
them above $10^8$ M$_{\odot}$ (depending on IMF
assumptions), with a gravitational magnification ranging between
1 and 3 magnitudes, the depth needed is of the order of
$J\sim$ 25.5, $H\sim$ 24.5 and $K\sim$ 24.0.

   The main emission lines expected in the near-IR domain, which
should allow an accurate redshift determination, are HeII
$\lambda$1640 and \lya. \lya \ can be detected
on near-IR spectroscopic surveys with 8-10~m class telescopes,
with a reasonable S/N over the redshift intervals z $\sim$ 7 to 18,
with some gaps depending on the spectral resolution (OH subtraction),
the atmospheric transmission, the intrinsic properties of
sources and the intergalactic medium (IGM) transmission. 
A joint detection with HeII
$\lambda$1640 should be possible for $z \sim$  5.5-7.0 (\lya \ 
in the optical domain), $z \sim$ 7--14 with both lines
in the near-IR. The typical expected fluxes for these
lines in our models range between $10^{-17}$ and a
few $10^{-18}$ erg/s/cm2. 
Thus, a reasonable S/N$\sim 3-5$ could be obtained
with a mid-resolution near-IR spectrograph such as ISAAC/VLT, with
exposure times of the order of a few hours (see also simulations by
Barton et al. 2003).
The detection of both HeII $\lambda$1640 and
\lya \  should allow to constrain the upper end of the IMF and the
age of these systems.

\section{Feasibility studies with GTs: A pilot
  project with VLT/ISAAC}\label{VLTsurvey}

\subsection{Survey Definition}\label{Survey}

\begin{figure}
  \caption{Composite $RzH$ image of the core of the lensing
cluster A1835, showing the critical lines at $z=1.55$, and a close
view of the main arc structure as seen on the R/F702W HST image.
}\label{fig:A1835}
\end{figure}

   We were granted ISAAC/VLT time to develop a pilot survey aimed at
searching for $z \gtapprox$ 7 galaxies using GTs.  Galaxy candidates 
are selected from ultra-deep $JHKs$ images (+
FORS/z and ISAAC/$SZ$ when available) in the 
core of gravitational lensing clusters for which deep optical imaging 
is also available, including HST data. We apply the Lyman drop-out 
technique to deep optical images. The main spectral discontinuity 
at $z \gtapprox 6$ shortward of \lya \  is due to the large neutral
H column density in the IGM. In addition, galaxies are selected according
to Fig.~\ref{fig:cc_theoriques}: we require a fairly red $J-H$ or $z/SZ-J$
color (due to the discontinuity trough the $z$, $SZ$ or $J$ bands), and
a blue $H-Ks$ colour corresponding to an intrinsically blue UV
restframe spectrum. The detection in at least
two bands longward of Lyman $\alpha$  and the combination with
the above $H-Ks$ colour criterion allows us to avoid contamination
by cool stars. After the photometric selection of candidates, a 
spectroscopic follow up was carried out with ISAAC/VLT. Our survey
will be presented in details in a forthcoming paper
(Richard et al. 2004, in preparation). 

  Two lensing clusters were selected for this pilot
survey. Table~\ref{tab:photom} summarizes the main properties of the
near-IR photometric data. 

\begin{itemize}

\item AC114 ($\alpha$=22:58:48.26 $\delta$=-34:48:08.3 J2000, $z=0.312$)
is very well known lensing cluster showing many multiple images with
redshifts ranging between $z\sim1$ and 4 (Smail et al.\ 1995; Natarajan et
al.\ 1998; Campusano et al.\ 2001; Lemoine-Busserolle et al.\ 2003). 
In addition to the near-IR data summarized in Tab.~\ref{tab:photom},
we have used optical photometry as reported in Campusano et
al.\ (2001), including deep HST/WFPC2-F702W(R) images. 


\item A1835 ($\alpha$=14:01:02.08 $\delta$=+02:52:42.9 J2000, $z=0.252$)
is the most X-ray luminous cluster in the X-ray-Brightest Abell-type
Clusters of Galaxies ($XBACS$) and the ROSAT Brightest Cluster samples
(Ebeling et al. 1998). A velocity dispersion of 1500 $km/s$ has been
obtained by Czoske et l.\ (2004) for this massive cluster from 
a spectroscopic survey with VIMOS/VLT of $\sim600$ cluster members up
to $R \le 23$. Strong and weak lensing were reported in this cluster
(Smail et al. 99, Limousin et al., in preparation). In addition to the
near-IR data summarized in Tab.~\ref{tab:photom}, we have obtained
deep $VRI$ images at CFHT, and HST/WFPC2-F702W(R) data (limitig
mag. R$_{702W}$ =28.9). A mass model is available mainly based on a
set of multiple images at z=1.56 (see Fig.~\ref{fig:A1835}). 

\end{itemize}
\begin{figure}
\includegraphics[width=4.0in,angle=270]{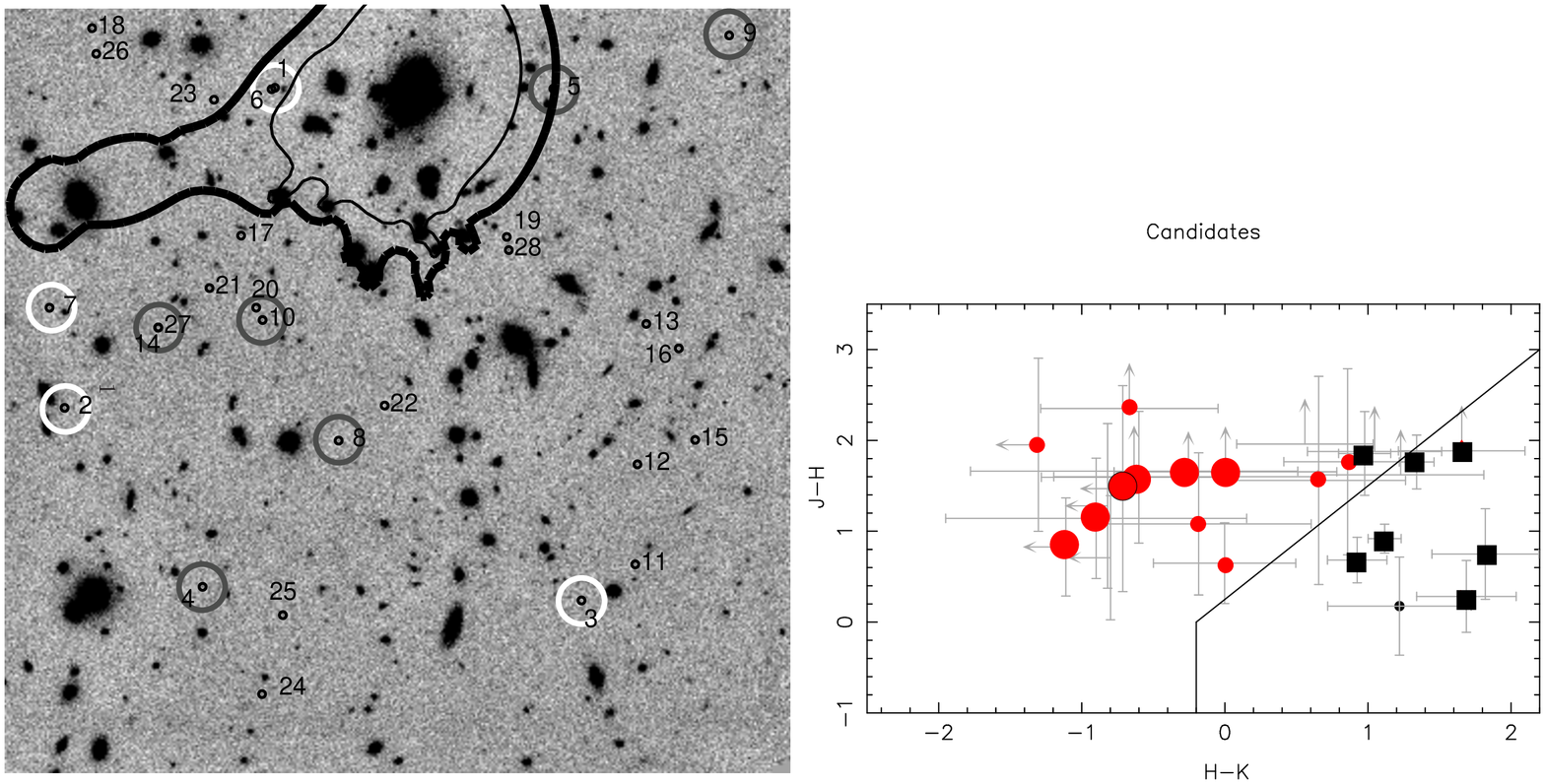}
  \caption{{\bf Left.} $H$-band image for A1835 showing the location
of the critical lines at z=1.5 (dashed curve) and z=10 (solid curve). The final
candidates are circled. Large (grey) solid circles display the first priority
candidates, whereas white circles correspond to the secondary targets
with a degenerate low-z solution in their redshift probability
distribution. The other points are faint secondary targets selected
only through their location on the color-color diagrams.
{\bf Right. }
Color-color diagram showing the location of all optical-dropouts
detected in A1835. The region selected corresponds to very high
redshift candidates at $z \sim$ 8--11. Optical dropouts fulfilling the
EROs definition are shown by black squares. Large and small red dots
display respectively the first priority candidates for spectroscopy
and the secondary targets for which a degenerate low-z solution exists
according to their redshift probability distribution.
}\label{fig:candidates}
\end{figure}

\begin{table}\def~{\hphantom{0}}
  \begin{center}
  \caption{Summary of Near-IR photometry obtained for our 2 lensing
clusters in this project: filter, filter effective wavelength, AB
correction ($m_{AB}=m_{Vega}+C_{AB}$) and, for each cluster, 
total exposure time, average seeing
measured on the original images, limiting (Vega)
magnitudes ($1 \sigma$ within a 1.5'' diameter aperture).}
  \label{tab:photom}
  \begin{tabular}{lcccccccc}\hline
       &                &          & AC114&        &        & A1835&        &       \\
\hline
Filter & $\lambda_{eff}$ & $C_{AB}$ & texp & seeing & m$_{lim}$& texp
& seeing & m$_{lim}$ \\
       &   (nm)& (mag) & (ksec) & (") & (mag) &(ksec) & (") & (mag) \\
\hline
$z$  & 919& 0.554 & - & - & - & 6.36 & 0.70 & 26.7  \\
$SZ$ &1063& 0.691 & - & - & - & 21.96 &  0.54 & 26.8 \\
$J$  &1259& 0.945 & 6.48 &0.52 & 26.0 & 6.48 & 0.65 & 25.6 \\
$H$  &1656& 1.412 &13.86 & 0.40 & 24.9 & 13.86 & 0.50 & 24.7 \\
$Ks$ &2167& 1.873 &18.99 & 0.34 & 24.8 & 18.99 & 0.38 & 24.7 \\
\hline
  \end{tabular}
 \end{center}
\end{table}

\subsection{Photometric selection of candidates}\label{selection}

   To look for high-$z$ galaxy candidates we have applied the
traditional Lyman break or drop-out technique. Optical/possibly $J$-band
drop-out and a blue UV rest-frame SED (i.e.\ peculiar $J-H$ and/or $H-K$
colors, as shown in Fig.~\ref{fig:cc_theoriques}),
allow us to select galaxy candidates with redshifts $z \sim$ 7--10
with intense ongoing star-formation. For the brightest candidates, we
have computed photometric redshifts based on the SED fitting of the
available photometric data using an adapted version of the code {\it
Hyperz} (Bolzonella et al. 2000).

Our present photometric survey in AC114 and A1835 with ISAAC/ESO
reaches an {\bf effective} depth after magnification correction by 1
magnitude of
26.4, 26.0, and 26.5 (3 $\sigma$, AB mags, 1.5'' aperture) in $J$, $H$, and $Ks$
respectively, and sufficient accuracy (typically $\ltapprox \pm 0.5$ mag
in $J-H$ and $H-K$) to apply our photometric selection criteria.
This is also illustrated in Fig.~\ref{fig:candidates}.
With a magnification $\mu >$ 1 mag over the entire field of view, 
the depth of our current observations is actually quite comparable to
or even deeper, close to the critical lines, than the NICMOS-UDF.
However, as the UDF, the {\bf effective} field of view surveyed in the
two clusters, after correction for lensing, remains relatively small,
$\sim 80\%$ of the UDF.

\subsection{VLT/ISAAC spectroscopic follow-up}\label{spectro}

\begin{figure}
  \caption{Sky-subtracted 2D spectra showing some of the emission
lines detected in our sample of high-$z$ candidates in A1835 and AC114. 
From top to bottom:
AC114-499 (z=7.17 candidate, Ly$\alpha$); A1835-1055 (z=7.89
candidate, Ly$\alpha$); A1835-775
(low-z contamination at z=1.89, double line [OII]3727) 
and A1835-2582 (low-z contamination at
z=1.67; H$\beta$, [OIII]4959,5007; Richard et al. 2003)}\label{fig:spectra}
\end{figure}

To look for faint emission lines, we have systematically explored
the 0.9-1.40 $\mu$ m domain (SZ and $J$ bands of ISAAC), where
\lya\ should be located for objects within the $7 < z < 10.5$ redshift
interval. Priorities are set according to the photometric redshift
probability distribution. First priority objects determine the observation
sequence. 
Each band has $\sim 600$ \AA\ width. 
With a spectral resolution for the sky lines of
$R=3100$, corresponding to the instrumental 1 arcsec slit width, 
the fraction of spectral band lost because of strong OH sky emission
lines, is of the order of 30\%.
Spectroscopic data were reduced using IRAF procedures and conforming to the ISAAC
Data Reduction Guide 1.5\footnote{{\tt
    http://www.hq.eso.org/instruments/isaac/index.html}}, using the
same procedure described in Richard et al. (2003).
Figure ~\ref{fig:spectra} displays 2D sky-subtracted spectra showing
some of the emission lines detected in our sample of high-$z$ candidates
in A1835 and AC114.  Typical line intensities range between $10^{-17}$ and a
few $10^{-18}$ erg/s/cm2.

\subsubsection{AC114}

  Two first priority candidates were observed in this cluster during a
single run, both of them very close to the z$\ge$7 critical lines,
with photometric redshifts ranging between 6.8 and 8.5. One of
them, AC114-499, displays a faint emission line at 9935 \AA \ compatible
with $z=7.17$ (if \lya, see Fig.~\ref{fig:spectra}). A solution at
$z=0.98$ (if this line corresponds to [OIII]$\lambda$5007) cannot be
excluded. This source has to be confirmed. 

\subsubsection{A1835}

  Two observing runs were conducted in this cluster, with the
following results, in addition to the first source likely to be at
$z=10.0$ (A1835-1916), which is discussed in the next
section: 

\begin{itemize}

\item A1835-1055: This source exhibits a unique emission line within
the wavelength interval explored, which could be $z=7.89$ (if \lya, see
Fig.~\ref{fig:spectra}). Given the spectral resolution of our
observations, it is unlikely [OII]$\lambda$3727 at $z=1.9$, because the doublet
would easily be resolved. It is difficult to reconcile with
[OIII]$\lambda$5007 at $z=1.16$, because [OIII]$\lambda$4959 is not
detected, or with H$\beta$ at $z=1.22$, because in this case
[OIII]$\lambda$5007 should be present. 

\item A1835-2582: This is a rather unusual emission line galaxy at
$z=1.68$, already studied in details by our group using these data 
(see Richard et al. 2003). This galaxy is extremely faint 
($M_B \sim$ --16.4), with a gravitational magnification of $\sim 2$ magnitudes.
Three lines, [OIII]$\lambda\lambda$4959,5007
and H$\beta$, were detected in our spectra. 

\item A1835-775: A double emission line is detected at the position of this
source. It is likely the [OII]$\lambda$3727 doublet at z=1.888.

\item A1835-1143: A faint line is detected at $\sim 3 \sigma$, to be
confirmed. 

\end{itemize}

   In addition to these sources, two other candidates observed in this
cluster did not show emission lines within the $J$ band. 
Of high priority, confirmation observations of A1835-1055 and A1835-1143
are scheduled.

\subsubsection{Results: A1835-1916}\label{1916}

\begin{figure}
  \caption{Thumbnail images of $16'' \times 22''$ around
the source \#1916 in A1835, obtained with ISAAC/VLT. From left to
right: original $H$-band image obtained
in February 2003 (t$_{exp}=$3.9h); $SZ$-band image ($1.06 \mu m$,
t$_{exp}=$6.1h), where the source was re-detected in May 2004; and
composite image in $SZ+H+Ks$, i.e. all the images where the source was
detected with S/N $\ge$ 3. Error bars correspond to mock simulations,
and magnitudes are computed within 1.5'' diameter aperture (the size
of the circle in these images). Given the time scale and the detection
levels, a TNO or a spurious detection seems highly unlikely.
}\label{fig:a1916}
\end{figure}

We recently obtained the first likely spectroscopic confirmation of a
$z=10.0$ galaxy in our sample (Pell\'o et al. 2004). This galaxy
(called A1835-1916) was one of the best candidates in the A1835 field, and
a good example of the search procedure. With our
original data the photometric
redshift probability distribution showed a clear maximum at redshift
$z_{\rm phot} \sim$ 9--11. This estimate was mainly corroborated
by a strong break of $\gtapprox$ 3.1 to 3.7 AB mags between
$VRI$ and $H$, which has a high significance independently of the definition
used for the limiting magnitudes. Our ISAAC/VLT spectroscopic
observations (29 June to 3 July 2003), under excellent seeing
conditions, resulted in the detection of one weak emission line at the
4-5 $\sigma$ level with an integrated flux of $(4.1 \pm 0.5)\times 
10^{-18}$ erg cm$^{-2}$ s$^{-1}$ at a wavelength of 1.33745 $\mu$m,
which appears on 2 different overlapping wavelength settings. When
identified as \lya\, in good agreement with the photometric SED,
we obtained $z=10.0$ for this source, the
most likely one given the data set available at this epoch. Its
properties have been widely discussed in different papers (see also
D. Schaerer's talk, this conference).  Also, our recent spectroscopic
observations in the $H$ band (1.6915 to 1.8196 $\mu$m, 2 overlapping
bands), display no other emission lines. In particular, neither
HeII$\lambda$1640 nor CIV$\lambda$1550 have been detected. We also exclude
all the likely solutions at $ z \sim2-2.6$, as well as most of
the $z \le 2$ possibilities.

The field around A1835-1916 has been reobserved
between 30 May and 6 June 2004
by Bremer et al.\ (2004) with NIRI/GEMINI in the $H$ band.
Surprisingly, the object is not seen anymore in these images, which
are at least $\sim$ 0.5 mag deeper than the ISAAC images taken 
approx.\ 15 months earlier.
The reality of our initial photometric detections is not questioned
(Bremer et al.\ reconfirm it using our data), and the detection
cannot be spurious, as jointly detected in 2 filters $HKs$
with S/N$\gtapprox4$ and 3 respectively --- according to our mock simulations
in a 1.5'' aperture. Bremer et al. conclude on the possible low-$z$
nature for this object. However, 
the analysis of new $SZ$ ($1.06 \mu m$) images
obtained 19 April ($\sim 4h$ exposure, 77 images) and 15 May ($\sim 2h$
exposure, 45 images) 2004, has provided some new puzzling results.
Although the two series of images have identical seeing and photometric
zeropoint, A1835-1916 is virtually non-detected or very faint in the 19 April
combination (S/N$\ltapprox$1.5-2), whereas it is clearly detected in the
final composition (see Fig.~\ref{fig:a1916}). The difference between 
the 2 series of images is about 0.5 magnitudes, but this value is still to be
taken with care. However, when we consider this result together with
our previous findings (the source was virtually non-detected in our $J$
images), and the recent non-detection by Bremer et al. in the $H$ band,
it seems quite clear that this source could be intrinsically
variable. Its nature (and hence also its redshift) presents quite a puzzle.

Possible explanations for faint transient/variable objects include
TNO, but the typical motions of $\sim 1'' /h$ are difficult to reconcile
with the time scales and rejection schemes. Alternatively this could be
a distant lensed SN (e.g.\ Marri \& Ferrara 1998;
Marri et al.\ 2000) or a tidal disruption of a star by a BH (cf.\ Stern et
al.\ 2004), but such events are thought to be very rare, and the
source show up at least twice in a 15 months period. The
UV variability of a distant quasar (e.g. Czerny, 2004;
Rengstorf et al., 2004) is a more likely explanation, given the
amplitude and time scale of the variability. In this case, the
emission line previously detected could correspond to the UV 
line of a high-$z$ AGN. 
The observed SED of the source is likely to be contaminated by
intrinsic variability.  

\subsubsection{Results: Confirmation Rate}\label{spectro_results}

\begin{figure}
\includegraphics[width=3.0in,angle=270]{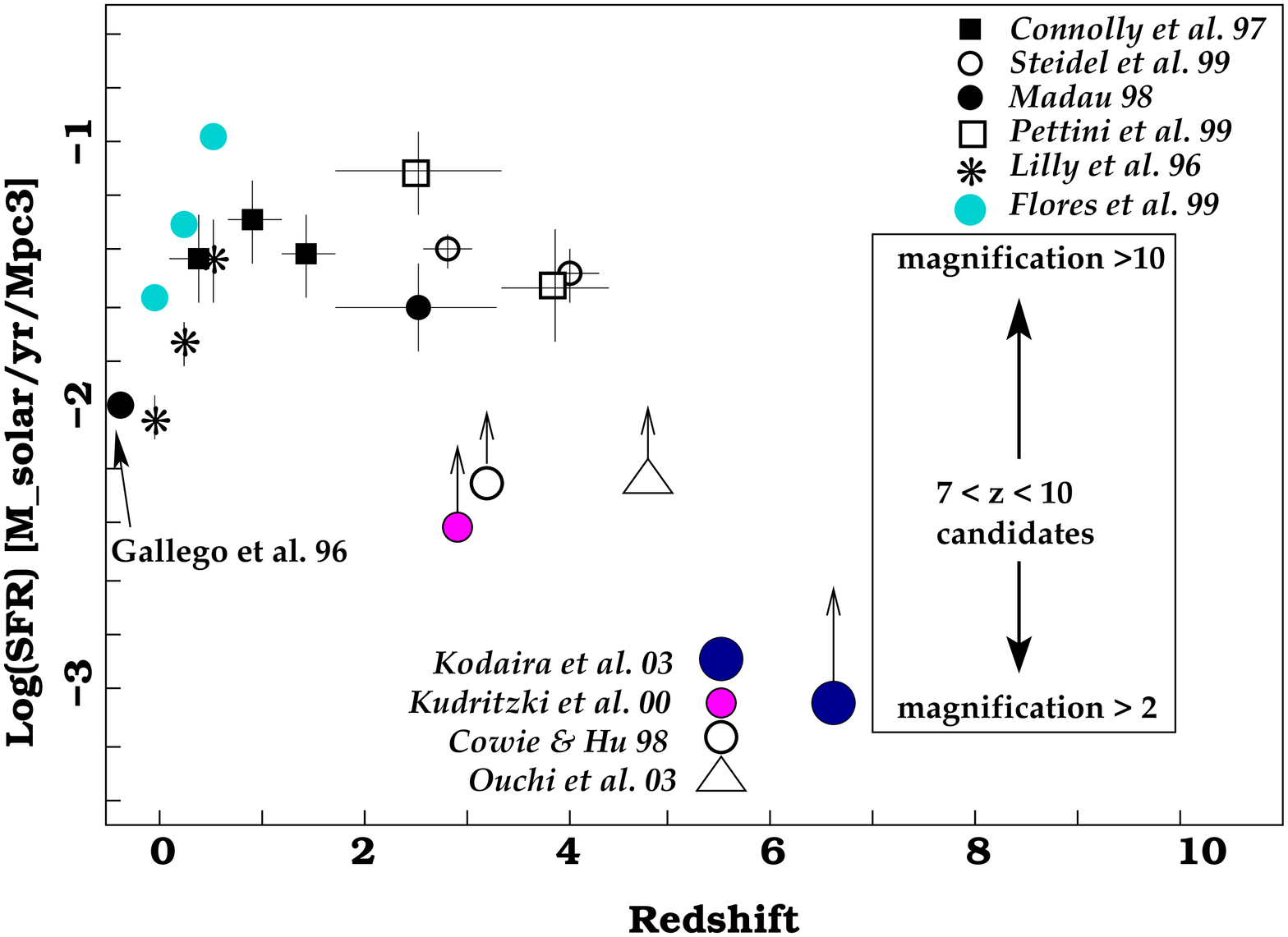}
  \caption{SFR density estimated from our
5 best candidates at 7 $\le$ z $\le$ 10
in the lensing cluster A1835, compared to previous values up to
$z \le 7$. This is a rough estimate, with upper and lower limits
corresponding to extreme
assumptions for the typical magnification factors ranging between
$\ge$ 2 and $\ge$ 10. This figure illustrates the inherent
uncertainties associated to such small-numbers statistics. 
}\label{fig:SFR}
\end{figure}

Our spectroscopic survey with ISAAC has targeted 2 out of $\sim$5 good
candidates in AC114, and 7 in A1835 (4 ``first priority'' targets and 3
secondary ones). From this sample of 9 targets, 2/3 of the objects
observed display emission lines. Actually 5 sources have clear
emission lines detected, and another one is still to be
confirmed. These emission-line sources can be classified as follows:

\begin{itemize}

\item {\bf High-$z$ $z\ge 7$ candidates}. This sample includes
A1835-1916, AC114-499, A1835-1055 and A1835-1143 (faint line still to be
confirmed). 

\item {\bf Low-$z$ sources}. This sample includes extremely faint
sources at z$\sim$1.5-2, such as A1835-2582 (z=1.68, Richard et
al. 2003) and A1835-775 (z=1.89). 

\end{itemize}

In summary, from 6 first priority targets observed in 2 clusters, we have clearly
confirmed 1 candidate (A1835-1916), which is found to be a puzzling
source, 2 are still to be confirmed at $z\ge7$, 1 is found to be a
low-$z$ contamination, and 2 of them do not show emission lines. From
the 3 secondary targets observed, only one is a possible $z\ge7$
source, whereas the other one showing emission lines is a faint low-$z$
galaxy. According to these numbers, the efficiency of our survey nowadays
could range between $\sim 30$ and 50\%, with interesting low-$z$
by-products. The final confirmation rate
will depend on the results obtained from our last observing run, which
is still uncomplete when writing this paper.
Improving the candidate confirmation, and possibly finding other
emission lines such as HeII$\lambda$1640 or
CIV$\lambda$1550 is of great interest to quantify the
nature/metallicity of these objects.
       
\section{Discussion}\label{discussion}

\subsection{Preliminary results on the 7$\le$ z $\le$ 10 photometric
  candidates}\label{candidates}

Applying the above broad-band selection criteria has provided
5-10 $z > 7$ galaxy candidates per lensing
cluster in the observed fields. We have used the lensing models to
derive the {\bf effective} areas and corresponding volumes surveyed at
the different source planes. Typical magnification factors range
between $\gtapprox$ 2 and $\gtapprox$ 10. The corresponding number density of
objects within $7 \ltapprox z \ltapprox$ 10 ranges between 
$2 \times 10^{-2}$ and $4 \times$ 10$^{-4}$ per
Mpc$^{-3}$, i.e. typically a few 10$^{3}$ objects deg$^{-2}$.
The main uncertainties are due to simplifying assumptions,
incompleteness corrections, and the values adopted for the typical
magnification factors of our sample. These quantities have been
evaluated with care and will be presented in a forthcoming paper
(Richard et al. 2004, in preparation). 

From the observed $H$ magnitudes and magnification factors derived
from the lens model, the typical star formation rates for these
candidates are estimated to be
a few \msun yr$^{-1}$ (lensing corrected, assuming a ``standard''
Salpeter IMF from 1--100 \msun). We have used simulations to correct
for observational incompleteness in our photometric data.
Fig.~\ref{fig:SFR} displays the star formation rate density derived 
from our 5 best candidates at 7 $\le$ z $\le$ 10
in the lensing cluster A1835, compared to previous values from the
literature. This figure illustrates the inherent
uncertainties associated to such small-numbers statistics. 
In short, a more detailed analysis and additional
observations are required to draw more firm conclusions
on the behaviour of the SFR density at $z>7$.

\begin{figure}
\includegraphics[width=3.5in,angle=0]{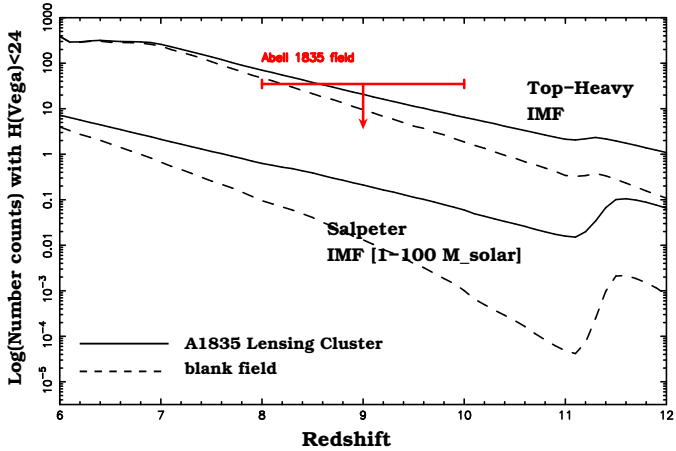}
  \caption{Comparison between the expected number counts of
galaxies in the field of ISAAC, up to $H\le$24, per redshift bin
$\Delta z=1$, in a blank field and in the field of the
strong lensing cluster A1835. Two different extreme assumptions for the IMF are
used: a Salpeter IMF, and a top-heavy IMF (stars forming
between 50 and 500 $M_{\odot}$). We display 
our own candidates in this field, corrected for photometric
completeness.
}\label{fig:counts}
\end{figure}

\subsection{Future developments}\label{future}

An important issue of this
project in view of the future facilities (such as EMIR at GTC and the
future KMOS for the VLT 2nd generation) is the strategy for target
selection, and the efficiency achieved in spectroscopic studies.
As shown in Sects.~\ref{z7-10} and ~\ref{VLTsurvey}, starbursts
could be detected from deep near-IR photometry based
on a measurement of two colors with accuracies of the order of ~0.3-0.5
mag. As shown in Fig.~\ref{fig:counts}, GTs
remain the ideal fields for the first prospective studies.
In this figure we compare 
the expected number counts of high-$z$ galaxies with $H\le$ 24, per
redshift bin $\Delta z=1$, in a blank field and in the field of the
strong lensing cluster A1835, for 2 different extreme assumptions for
the IMF: a standard Salpeter IMF, and a top-heavy IMF (stars forming
between 50 and 500 $M_{\odot}$). A simple
Press-Schechter formalism for the abundance of halos
and standard $\Lambda$CDM cosmology were considered in these
order of magnitude simulations, as well as a conservative
fixed fraction of 10\% of the baryonic mass in halos converted into stars.
Strong lensing fields are a factor of
$\sim 5-10$ more efficient than blank fields of the same size in the z
$\sim 7-11 $ domain, all the other conditions beeing the same. In
addition, the large magnification factors achieved on unresolved
sources facilitates the spectroscopic follow up from ground-based
telescopes. Additional discussion can be found in Pell\'o et
al. (2004b). 

  Large Surveys devoted to the identification and study of $z\ge7$
galaxies are already planned. It is crucial to increase the
multiplexing capabilities of near-IR spectrographs, as well as the
field of view and the typical wavelength interval which can be
explored in a single shot. 
An important project in this area is the
GOYA Survey, using the new multi-object near-IR spectrograph EMIR at
GTC, starting in 2006 (Garz\'on et al. 2003).

\section{Conclusions}\label{conclusions}

   The bottom line in the present searches for z $\sim 7-11 $
galaxies is efficiency. The different techniques used to identify
high-$z$ candidates (narrow-band or broad-band searches) are
complementary. They have to be improved and optimized in the
near future to enhance the confirmation rates. Both the 
identification procedures and the individual 
high-$z$ sources issued from these procedures should be validated, if
possible through complementary observations of the same field. 

The search efficiency should be significantly improved
by the future near-IR multi-object ground-based and space
facilities. However, strong lensing clusters remain a factor of $\sim
5-10$ more efficient than blank fields in the z $\sim 7-11 $ domain,
within the FOV of a few arcminutes around the cluster core,
for the typical depth required for these survey projects.

\begin{acknowledgments}

We are grateful to A. Ferrara, M. Lemoine-Busserolle, D. Valls-Gabaud,
G. Mathez, T. Contini and F. Courbin for comments and discussions.
Based on observations collected at the European Southern
Observatory, Chile (70.A-0355, DDT 271.A-5013, 71.A-0397, 73.A-0471), 
the NASA/ESA Hubble Space Telescope
operated by the Association of Universities for Research in Astronomy, Inc.,
and the Canada-France-Hawaii Telescope operated by the
National Research Council of Canada, the French CNRS
and the University of Hawaii.
Part of this work was supported by the CNRS
and the Swiss National Foundation.

\end{acknowledgments}

\end{document}